\documentclass[natbib]{svjour3}                     
\smartqed  
\usepackage{graphicx}

\usepackage{mathptmx}      
\journalname{Space Science Reviews}

\begin{document}

\title{The Puzzle of the Flyby Anomaly
}


\author{Slava G. Turyshev \and Viktor T. Toth}


\institute{Slava G. Turyshev \at
           Jet Propulsion Laboratory,
       California Institute of Technology,\\
       4800 Oak Grove Drive, Pasadena, CA 91109 USA \\
              \email{turyshev@jpl.nasa.gov}           
\and
           Viktor T. Toth \at
           Ottawa, ON  K1N 9H5, Canada\\
              \email{vttoth@vttoth.com}
}

\date{Received: date / Accepted: date}

\maketitle

\begin{abstract}
Close planetary flybys are frequently employed as a technique to place spacecraft on extreme solar system trajectories that would otherwise require much larger booster vehicles or may not even be feasible when relying solely on chemical propulsion. The theoretical description of the flybys, referred to as gravity assists, is well established. However, there seems to be a lack of understanding of the physical processes occurring during these dynamical events. Radio-metric tracking data received from a number of spacecraft that experienced an Earth gravity assist indicate the presence of an unexpected energy change that happened during the flyby and cannot be explained by the standard methods of modern astrodynamics. This puzzling behavior of several spacecraft has become known as the flyby anomaly. We present the summary of the recent anomalous observations and discuss possible ways to resolve this puzzle.

\keywords{Flyby anomaly \and gravitational experiments \and spacecraft navigation.}
\end{abstract}

\section{Introduction}

Significant changes to a spacecraft's trajectory require a substantial mass of propellant. In particular, placing a spacecraft on a highly elliptical or hyperbolic orbit, such as the orbit required for an encounter with another planet, requires the use of a large booster vehicle, substantially increasing mission costs. An alternative approach is to utilize a gravitational assist from an intermediate planet that can change the direction of the velocity vector. Although such an indirect trajectory can increase the duration of the cruise phase of a mission, the technique nevertheless allowed several interplanetary spacecraft to reach their target destinations economically \citep{Anderson1997,VanAllen2003}.

Notable missions\footnote{Several additional missions have used planetary assists to reach their target destinations, including Mariner 10 (Venus and Mercury), Pioneer 10 and 11 (Jupiter and Saturn), and also Voyager 1 and 2, which used gravity assists from Jupiter to reach Saturn. Voyager 2 continued to Uranus and Neptune, using the gravity assist of each planetary encounter to target the spacecraft to the next planet. The most feasible plans of space missions inward toward the Sun (such as the Ulysses mission that used a Jupiter flyby to form a trajectory outside the ecliptic plane) and outward to Pluto (such as the New Horizons mission that used a Jupiter flyby to increase significantly the craft's velocity) depend on gravitational assists from Jupiter.} that relied on an Earth gravity assist maneuver and are relevant to the main topic of this paper include Galileo\footnote{\tt http://www2.jpl.nasa.gov/galileo/}, which had two encounters with the Earth and one each with Venus and an asteroid to reach Jupiter more quickly; the Near Earth Asteroid Rendezvous\footnote{\tt http://near.jhuapl.edu/} (NEAR Shoemaker) mission;
the Cassini mission\footnote{\tt http://saturn.jpl.nasa.gov/} with encounters with Venus, Earth, and Jupiter to speed it on its way to Saturn; and the European Space Agency's Rosetta mission\footnote{\tt http://sci.esa.int/science-e/www/area/index.cfm?fareaid=13} en route to an encounter with the comet 67 P/Churyumov-Gerasimenko.

However, during the Earth flybys, these missions experienced an unexpected navigational anomaly. In the following, we discuss the nature of gravity assist maneuvers, characterize the flyby anomalies experienced by these spacecraft, and discuss the challenges that one faces in attempting to find an explanation of this effect.

\section{Gravity assist maneuvers}

A gravity assist maneuver is a specific application of the restricted three body problem, in which an effectively massless test particle (such as a spacecraft) moves in the combined gravitational field of two larger bodies. When the larger bodies move in circular orbits, the problem is known as the circular restricted three-body problem, or Euler's three-body problem \citep{Euler}, among other names. In this problem, the energy and momentum of the test particle are not conserved, although other conserved quantities exist. The energy gain or loss by the test particle is offset by a corresponding loss (gain) in energy by the two larger bodies in the system, however, due to the differences in mass, the corresponding changes in the larger bodies' velocities are not perceptible.

The circular restricted three-body problem is exactly solvable: after a suitable set of generalized coordinates are chosen, the solution can be expressed in the form of elliptic integrals \citep{Whittaker}. The problem has also been analyzed by use of the method of patched conics \citep{Breakwell-etal-1961,Battin-1987}, which notionally patches two conics together at the trajectory's intersection with the sphere of influence surrounding the smaller mass. As well, the problem can be addressed by numerically integrating the equations of motion using a suitably chosen integration method of sufficient accuracy.

However, precision calculation of the trajectory of a spacecraft in the vicinity of a planet requires detailed analysis that takes into account all the effects including a complicated gravitational potential (usually represented in the form of spherical harmonics), perturbations due to the gravitational influence of the planet's moons, if any, the pressure of light and thermal radiation received from the Sun and the planet, drag forces that may be present in the planet's upper atmosphere, and on-board events such as thruster firings. The precision with which a trajectory can be computed is determined by the accuracy with which all these effects can be accounted for. When sufficient accuracy is achieved, this results in a diminishing difference between computed an observed values of tracking observables (e.g., radio-metric Doppler and ranging data.) The methods and tools used for high-precision navigation have seen major improvements \citep{Moyer-2003}.

In particular, when a spacecraft receives a gravity assist from the Earth, its trajectory must be computed after taking into account the gravitational effects of the Earth and Moon, and nongravitational effects in the near Earth environment, including the upper atmosphere, thermal recoil forces generated on-board due to heat dissipation processes, etc. The dynamics of an Earth flyby are well understood, and are a subject of continuous study thanks to the large number of Earth-orbiting satellites, in particular as a result of the enormous success of the GRACE mission, which led to the construction of the GRACE Gravity Model\footnote{For details on the GRACE mission and the GRACE Gravity Model 02 (GGM02), please consult \tt http://www.csr.utexas.edu/grace/gravity/} \citep{Tapley-etal-2005}.

Nonetheless, during the Earth flyby of several spacecraft, a small, anomalous increase in velocity was observed \citep{JPL2008}. Although these anomalies were not mission critical events and did not prevent the spacecraft to proceed to their ultimate destinations, the puzzle of the flyby anomaly, as this effect became known, remains unresolved.

\section{The Earth flyby anomaly}

\begin{figure}[t!]
\begin{center}
\includegraphics[width=0.85\linewidth]{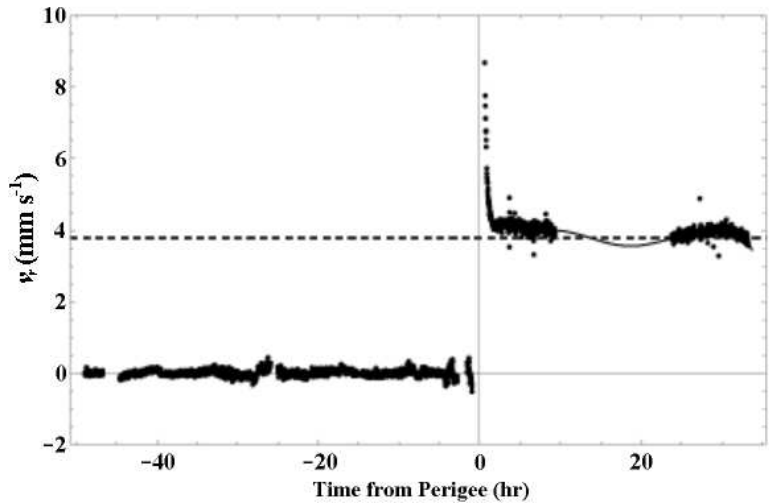}
\end{center}
\caption{Doppler residuals (observed minus computed Doppler frequency, converted into units of line-of-sight velocity) during Galileo's December 8, 1990 Earth flyby \citep{JPL2009}. Residuals are based on a pre-perigee fit to radio-metric Doppler data, and demonstrate the anomalous post-perigee velocity increase.}
\label{fig:GLLI}
\end{figure}

Several spacecraft, including Galileo, NEAR, Cassini, and Rosetta, utilized Earth flybys to achieve their desired trajectories. For all four of these flybys an anomalous change in the modeled flyby velocity is required \citep{Anderson-etal-2007} in order to fit the Deep Space Network (DSN) Doppler and ranging data. This increase can be represented by a fictitious trajectory maneuver at perigee, or it can be demonstrated by fitting the pre-encounter data, and subsequently using the resulting trajectory to predict the post-encounter data. The difference between the actual post-encounter data and the predicted data is consistent with the velocity change determined from the fictitious maneuver.

This anomalous trajectory behavior was first noted after the first Earth gravity assist for the Galileo spacecraft on 8 December 1990 (Figure~\ref{fig:GLLI}). The increase in the asymptotic velocity $V_\infty$ of its hyperbolic trajectory was $\Delta V_\infty=3.92\pm 0.08$~mm/s. The efforts by the Galileo navigation team and the Galileo radio science team to find a cause for this anomaly were unsuccessful. Two years later, during Galieo's second Earth flyby on 8 December 1992, the Tracking and Data Relay Satellite (TDRS) system was scheduled to track the spacecraft \citep{edwards1993}. This time the perigee altitude was lower, 303~km vs. 960 km. Consequently, any anomalous velocity increase was masked by atmospheric drag. However, results were published for the two Galileo flyby navigational anomalies \citep{edwards}.

After the Earth flyby by the NEAR spacecraft on 23 January 1998, at an altitude of 539 km, the anomalous velocity increase was observed once again, with a magnitude of $\Delta V_\infty=13.46\pm 0.13$~mm/s. Results were presented at a spaceflight conference \citep{Antreasian-Guinn-1998}, including a reanalysis of the two Galileo flybys, and with all three flybys based on the best Earth gravity field available in August 1998.

Subsequently, the Earth flybys of additional spacecraft were studied, in order to establish if an anomalous velocity change is present in the tracking data. When Cassini flew by the Earth on August 18, 1999 at an altitude of 1175~km, thruster maneuvers prevented the immediate detection of any anomalous behavior. However, subsequent analysis showed an anomalous velocity change of $2\pm 1$~mm/s in magnitude.

The navigation team of the European Space Agency's Rosetta spacecraft, analyzing the spacecraft's trajectory during its Earth flyby on 4 March 2005, also observed an anomalous change in that spacecraft's velocity \citep{Morley-Budnik-2006}, at $\Delta V_\infty=1.80\pm 0.03$~mm/s, as that spacecraft flew by the Earth at an altitude of 1956~km.

However, during the subsequent flyby by the MESSENGER spacecraft on 2 August 2005, at an altitude of 2347 km, no anomalous velocity change was seen.

Attempts to find this anomalous orbital energy increase for flybys of other planets have failed so far because the gravity fields are not known nearly as well as for the Earth. Perhaps in the future, models of the gravity fields of Venus and Mars, determined from orbiter data, could become sufficiently accurate that DSN Mars and Venus flyby data from earlier NASA missions could be used to search for an anomalous $\Delta$V$_\infty$. These old DSN data can be retrieved from NASA's National Space Science Data Center (NSSDC) archive, where they were deposited by various radio science teams.

\begin{table}[t!]
\begin{center}
\caption{Summary of Earth flyby spacecrafts, based on \citep{JPL2008}.}
\begin{tabular}{|p{2.5cm}|l|l|l|l|l|}\hline
Quantity&Galileo&NEAR&Cassini&Rosetta&Messenger\\\hline\hline
Speed at infinity, km/s &8.949  &6.851  &16.01  &3.863  &4.056\\
Minimal altitude, km    &956.053    &532.485    &1171.505   &1954.303   &2336.059\\
Spacecraft mass, kg &2497.1 &730.40 &4612.1 &2895.2 &1085.6\\
$\Delta V_\infty$, mm/s   &3.92$\pm$0.08  &13.46$\pm$0.13 &-  &1.82$\pm$0.05  &-\\
Gained energy, J/kg &35.1$\pm$0.7   &92.2$\pm$0.9   &-  &7.03$\pm$0.19  &-\\
Inclination to equator  &142.9$^\circ$  &108.8$^\circ$  &25.4$^\circ$   &144.9$^\circ$  &133.1$^\circ$\\
Deflection angle   &47.46$^\circ$  &66.92$^\circ$  &19.66$^\circ$  &99.396$^\circ$ &94.7$^\circ$\\\hline
\end{tabular}
\label{tb:earth-flybys}
\end{center}
\vskip -15pt
\end{table}

The results from Galileo, NEAR, Cassini, and Rosetta, as well as the null result from MESSENGER, summarized in Table \ref{tb:earth-flybys}, suggest that a relationship may exist between the magnitude of the flyby anomaly and the altitude and geometry of the flyby trajectory. Indeed, an empirical relationship was proposed by \cite{JPL2008}. While the reported anomalous energy changes were claimed to be consistent with a simple empirical formula that fits most of the data in Table \ref{tb:earth-flybys}, the physical origin of this expression is yet unclear. Furthermore, the data seem to indicate that the search for a phenomenological model is yet unfinished.

\section{Conclusions}

Multiple experiments performed by different teams using a variety of spacecraft have firmly established the existence of an anomalous effect during Earth flybys. The cause of this effect remains unknown. \cite{Lammerzahl} have considered, and dismissed, a variety of possible conventional causes, including effects due to the atmosphere, tides, electric charge and magnetic moment of the spacecraft, radiation pressure due to the Earth's albedo, solar wind, and spin-rotation coupling. \cite{Antreasian-Guinn-1998} have earlier dismissed explanations involving the Moon, the Sun, and relativistic effects.

Another possibility is the incorrect modeling of the flybys and their realization in navigational software. We must emphasize the need for a consistent treatment of the gravitational assist problem from the standpoint of the modern theory of astronomical reference frames. Of particular concern is the theoretical treatment and practical realization of the concept of spheres of influence in the software codes used to navigate the spacecraft \citep{Moyer-2003}. In this respect, an appropriate treatment must properly describe the dynamical and signal propagation effects relevant to the transition between various reference frames \citep{Soffel-etal-2003} in the overlapping region. For instance, in the case of the Earth flyby problem these frames are the solar system barycentric and geocentric frames. Appropriate values for astronomical constants must be used to model the spacecraft motion when it transits from the sphere of influence of one body to that of another. Finally, one must also make sure that all the ancillary data derived by external means (such as VLBI, lunar laser ranging, spacecraft radar etc.) and used to describe solid-Earth effects (namely Earth's precession, nutation, sidereal rotation, polar motion, tidal effects, tectonic plates drift, etc.) have consistent definitions within the overlapping region between the frames used to treat the flyby problem, which currently is not the case.

It is tempting to consider the flyby anomaly in conjunction with other spacecraft navigation anomalies, including the Pioneer anomaly \citep{JPL2002}, and the anomalous increases in the Astronomical Unit (AU) and the eccentricity of the Moon's orbit \citep{JPL2009}. However, there is no direct evidence that these anomalies are related, and at least in the case of the Pioneer anomaly, the possibility of a conventional explanation cannot be dismissed \citep{MDR2005,Toth:2009se,JPL2009}.

Nonetheless, the study of the Pioneer anomaly teaches a very valuable lesson concerning the preservation of old navigational and other spacecraft data. As discussed above, it may be possible to re-analyze radio tracking data from past Mars and Venus flybys once more accurate models of the gravitational fields of these planets become available. Such a re-analysis will only be possible if the archived tracking data is still available. So long as the possibility exists that the flyby anomaly has an origin in on-board systematics, it is also advisable to preserve the engineering telemetry record of the spacecraft in question.

The work outlined above is important but rather tedious. Yet it must be done before any discussion of new physics as an explanation for the flyby anomaly can seriously take place. To that extent, a continuing effort is needed to better characterize the anomaly, using both past and future flyby data, similar to the on-going efforts to study the Pioneer anomaly \citep{MDR2005}.  Some of this work has already begun, for example in the context of the International Flyby Collaboration\footnote{For more information on the International Flyby Collaboration at ISSI, please visit team's website at \tt http://www.issibern.ch/teams/investflyby/} at the International Space Science Institute (ISSI). However, significant additional effort is required. This paper intends to motivate initiation of such a work in the near future.

By 2009, the existence of the flyby anomaly has been confirmed by a variety of teams with several spacecraft. While it is unlikely that the origin of the flyby anomaly will prove to be anything other conventional in nature, because of its importance for high-precision navigation further investigations into the physics of this puzzling effect are certainly justified.

\begin{acknowledgements}
The work of SGT was carried out at the Jet Propulsion Laboratory, California Institute of Technology, under a contract with the National Aeronautics and Space Administration.
\end{acknowledgements}

\end{document}